\documentstyle[psfig]{laa}    
\begin{document}
\psfigurepath{figures/}						
\thesaurus{08(08.06.1, 08.16.5, 08.09.2 EX Lup)}
\title{The outburst of the T Tauri star EX Lupi in 1994}
\author{Thomas Lehmann\inst{1,2} \and Bo Reipurth\inst{1}
  \and Wolfgang Brandner \inst{1,3}} 
\offprints{Thomas Lehmann, e-mail: {\em lehmi@sol.astro.uni-jena.de}}
\institute{European Southern Observatory, Casilla 19001, Santiago 19, Chile
  \and Astrophysikalisches Institut und Universit{\"a}ts-Sternwarte Jena,
       Schillerg{\"a}{\ss}chen 2, D-07740 Jena, Germany
  \and Astronomisches Institut der Universit{\"a}t 
       W{\"u}rzburg, Am Hubland, D-97074 W{\"u}rzburg, Germany}
\date{-received date- ; -accepted date-}
\maketitle

\def\izq#1{\hbox to -1.5pt{\hss#1}}

\begin{abstract}
We have observed an outburst of the T Tauri star EX Lup in March 1994. We
present both photometric (BVR) and
spectroscopic (low and medium resolution) observations carried out during the
decline after outburst. The star appears much bluer during outburst due to
an increased emission of a hot continuum. This is accompanied by a strong
increase of the veiling of photospheric lines. We observe inverse P Cygni
profiles of many emission lines over a large brightness range of EX Lup.
We briefly discuss these features towards the model of magnetospherically
supported accretion of disk material.
\end{abstract}


\section{Introduction}
The variability of EX Lup was discovered by Miss E. Janssen in 1944 while
examining spectral plates at the Harvard Observatory (McLaughlin 1946).
Herbig (1950) first pointed out the similarity of EX Lupi's spectral 
characteristics and T Tauri stars with strong emission lines of H, CaII,
FeII, and HeI. In one of the spectrograms he obtained
in 1949/1950 the H and CaII lines clearly show an inverse P Cygni profile.
Herbig (1977a) assigned the spectral type of M0:eV using
the 5850-6700 {\AA} range.
Photographic and visual light-curves covering a century of
observations revealed the irregular photometric behaviour of the star
(McLaughlin 1946, Bateson et al. 1990). Outbursts of up to 5 magnitudes
may occur, but the star normally shows only small amplitude irregular
variations. The most prominent
events last about one year. The typical recurrence time scale of
outbursts is of the order of a decade.\\
Up to now there are only a few other stars known with comparable outburst
characteristics (Herbig 1989). This small group of very
active T Tauri stars has been called EXors or sometimes
sub-FUors. Both names point to an affinity
to the so called FU Orionis stars (FUors). FUors are another group of
young low mass stars with even stronger outbursts lasting
for decades. Unlike EXors, during an outburst FUor spectra turn from T Tauri
characteristics to that of much earlier F or G supergiants lacking strong line
emission (Herbig 1977b). FUors have high mass accretion rates
($\dot{M}_{acc} \geq 10^{-4} M_{\odot} yr^{-1}$,Hartmann 1991)
and strong winds (e.g. Calvet et al. 1993) and they
may be the source that drive Herbig-Haro flows (Reipurth 1989).\\
EXors are little studied, but potentially of great interest because they may
represent an intermediate level of activity between ordinary active T Tauri
stars and full blown FU Orionis eruptions. In order to cast further light on
this interpretation, we have followed some EXors spectroscopically and
photometrically during 1993 and 1994.
 

\section{Observations}
The star EX Lup has been at a low level of activity during the 1980's. In the
early 1990's this situation changed and the star became active
(Jones et al. 1993, Hughes et al. 1994).
Amateur observations (Variable Star Section of the Royal Astronomical Society
of New Zealand, unpublished) indicated a strong brightening in
February/March 1994. Patten (1994) reports some follow-up photometric and low
resolution spectroscopic observations of the same outburst.\\
In this paper we present part of our optical observations of EX Lup taken
at ESO, La Silla. We
concentrate on data obtained during the outburst in March 1994 and include
some spectroscopic observations carried out in August 1994 when the star only
exhibited post-outburst low level activity.
A complete presentation of our data will appear in a future paper.


\section{Photometric results}
Differential CCD photometry has been carried out at the 0.9m-Dutch and the 1.54m-Danish
telescopes. This photometry was later calibrated with respect to standard stars
including extinction and colour
corrections. All reductions have been made with the
APPHOT package in IRAF. Typical errors
(1$\sigma$) in the differential photometry are $\Delta$B=0.005,
$\Delta$V=0.004, $\Delta$R=0.004 whereas the absolute magnitude scale itself
is accurate to about 0.01 in all three colours.\\
The resulting lightcurves in B, V, and R are presented in Fig. 1.
The maximum occurred between February 25 and March 4 (Herbig, priv. comm.).
The fading tail of the eruption can be described as an exponential decline
with small fluctuations superimposed. Variability of more than 0.1mag is
present on timescales of less than one hour (e.g. March 6.3, see also Patten
1994).
Figure 2 displays the colour change in B-V during the decline. The star clearly
becomes redder when fading. For comparison we have included some points
close to minimum light taken from the literature. The outburst amplitude was
about
$\Delta$V=2.0mag and $\Delta$B=2.6mag~.
\begin{figure}[t]
  \centerline{\psfig{figure=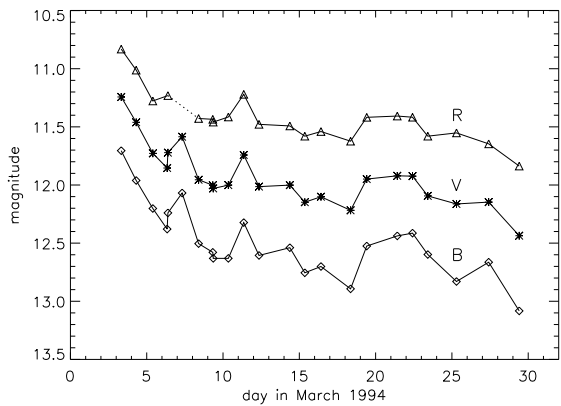,height=5.8cm}}		
  \caption[ ]{Photometry obtained in March 1994. R {\em (top)\,}, V {\em 
(middle)\,} and B {\em (bottom)\,} lightcurves.}
\end{figure}
\begin{figure}[t]
  \centerline{\psfig{figure=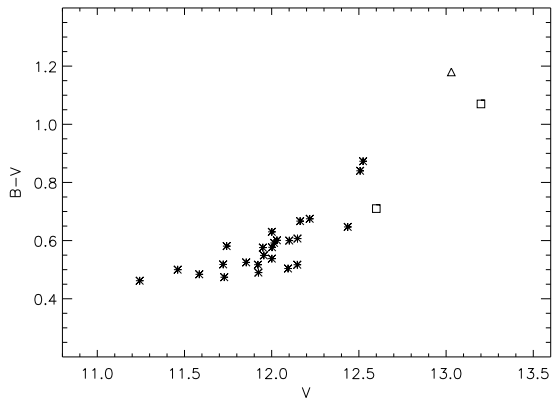,height=5.8cm}}		
  \caption[ ]{V -- (B-V) magnitude-colour diagram. A few measurements were 
taken from literature: Herbig et al.1992 {\em (squares)\,}, Bastian \&
Mundt 1979 {\em (triangle)\,}.}
\end{figure}


\section{Spectroscopic results}
Spectroscopic observations in the blue spectral range were carried out
during the first few nights in March 1994 on the ESO-1.52m telescope using the
Boller \& Chivens spectrograph at 1.2{\AA} resolution. After the decline of EX
Lup we obtained
post-outburst spectra in the same wavelength region at resolutions of 1.5{\AA}
and 12{\AA}
at the 3.5m-NTT with EMMI in August 1994.
All spectra have been reduced with the CTIOSLIT package in IRAF. Observations
of spectrophotometric standards and nightly extinction curves allowed
for a flux calibration.\\
In Fig.3 we present two spectra of EX Lup: one close to the outburst maximum
and the other at low activity almost half a year after the eruption.
Some of the emission lines of H, CaII, FeII, HeI, and HeII are indicated.
Under the assumption that the total light can be decomposed into an
underlying T Tauri star photosphere, a continuum source, and superimposed
emission lines, we now discuss the different spectral components and
their variability. 
\begin{figure*}[ht]
  \centerline{\hspace{-7mm} \psfig{figure=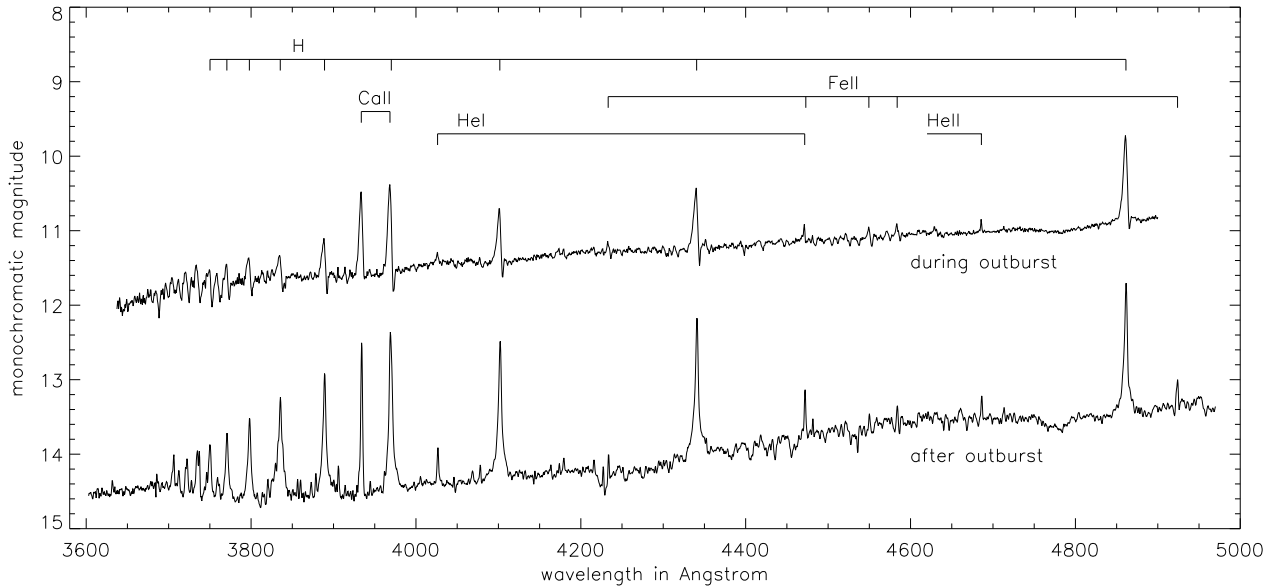,height=8.4cm}} 
  \caption[ ]{Medium resolution spectra of EX Lup. Fluxes are expressed in
magnitudes according mag = -2.5 log (F$_{\nu}$/F$_{0})$ where F$_{\nu}$ is the
flux per unit frequency and
F$_{0}=3.68*10^{-20}$ erg/cm$^{2}$/s/hz. Some emission lines of H, CaII, FeII,
HeI, and HeII are indicated.\ \
Inverse P Cygni profiles in the strongest emission lines
are visible in the outburst spectrum from March 3 {\em (top)\,}.\ \ 
Photospheric absorption features (e.g. CaI 4227) appear in the post-outburst
spectrum from August 16 {\em (bottom)\,}.}
\end{figure*}

\subsection{Continuum emission}
A powerful method to determine the continuum excess emission is to determine
the veiling by comparison with
spectra of stars of the same spectral type and luminosity class but lacking
any disk signature (Hartigan et al. 1989,
1991). The accuracy of the veiling determination decreases rapidly when the
emission component
exceeds the photospheric luminosity. In the case of EX Lup during its eruption
we therefore did not intend to derive the veiling and the true excess emission
spectrum by comparison with spectral type standards, but
we could examine the spectral variability caused by the outburst.\\
No photospheric absorption features are seen during the
outburst (upper spectrum in Fig.3) but they appear in the post-outburst
spectrum.
Thus the major source of variability presumably is a featureless continuum.
Therefore, a difference spectrum between outburst and post-outburst spectra
should be a good measure of the continuum emission spectrum.
In Fig. 4 we plot two difference spectra at low resolution. The first shows
the difference between an outburst (March 3) and a post-outburst (August 16)
spectrum, while the second shows the difference between two post-outburst
(August 18 and 16) spectra which displays normal low-level variability.
The continuum emission spectrum displaying the normal low-level activity
is bluer than the continuum emission present during outburst.

\subsection{Emission lines}
The most intriguing features in the spectra of EX Lup are strong emission
lines. The Balmer series can be seen up to H15 especially during times of minimum activity. Equivalent widths and fluxes of individual lines are given in
Table 1. Essentially all strong emission lines have increasing fluxes as the
star brightens. However due to the steep rise of the continuum the equivalent
widths decrease, which is also evident in the data from Patten (1994) at 
H$\alpha$, H$\beta$, and H$\gamma$ during the maximum.
Obviously the CaII lines have a larger flux amplification during the outburst
than the Balmer lines. There is some indication that line fluxes of metals
do not increase while the star goes into outburst (CaI, FeII, SrII).\\
\begin{figure}[t]
  \centerline{\psfig{figure=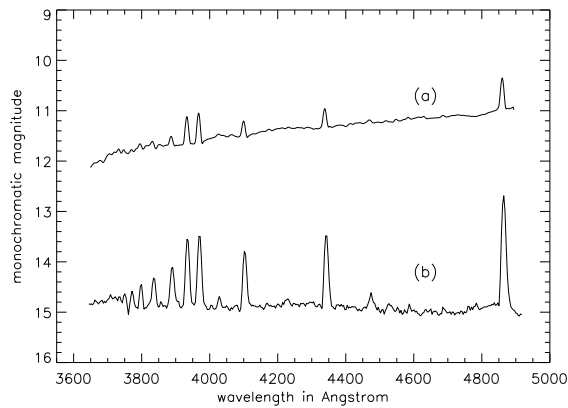,height=5.8cm}}		
  \caption[ ]{Excess emission - Comparison of outburst and low level
variability at low spectral resolution. These spectra are a good
approximation to the excess emission spectra which are superposed on the
stellar photospheric spectrum. \ \ {\bf (a)} The outburst as given
by the difference of the spectra from March 3 and August 16.
 \ \ {\bf (b)} Post-outburst variability derived from spectra from August 18
minus August 16. Note the Balmer jump emission and the bluer continuum.}
\end{figure}
\begin{table}[tp]
  \caption[ ]{Comparison of selected emission lines at different levels of
activity. Equivalent widths and line fluxes during the outburst
measured on March 3 ($^{\rm high}$) and in the post-outburst spectrum on August
16 ($^{\rm low}$)}
{\scriptsize
  \begin{center} 
    \begin{tabular}{lrrrr} \hline \hline
 & & & & \\[-1ex]
Identification & W$_{\lambda}^{\rm high}$ & W$_{\lambda}^{\rm low}$ & flux$^{\rm high}$ & flux$^{\rm low}$ \hspace{0.1em} \\
 & {\AA}\hspace{1em}  & {\AA}\hspace{1em}  & $\frac{10^{-14}{\rm erg}}{{\rm cm}^{2}{\rm s}\,{\rm \AA}}$ & $\frac{10^{-14}{\rm erg}}{{\rm cm}^{2}{\rm s}\,{\rm \AA}}$ \\
 & & & & \\[-1ex] \hline
 & & & & \\[-1ex]
H11 3771	& -1.0 \hspace{0.8em} &  -4.2 \hspace{0.1em} &  16 \hspace{1.0em} &  6 \hspace{1.5em} \\
H10 3798	& -1.5 \hspace{0.8em} &  -7.0 \hspace{0.1em} &  25 \hspace{1.0em} &  9 \hspace{1.5em} \\
H 9 3835	& -1.2 \hspace{0.8em} & -12.1 \hspace{0.1em} &  20 \hspace{1.0em} & 16 \hspace{1.5em} \\
H 8 3889	& -2.7 \hspace{0.8em} & -13.0 \hspace{0.1em} &  44 \hspace{1.0em} & 18 \hspace{1.5em} \\
SiI 3906	& -0.2 \hspace{0.8em} &  -0.8 \hspace{0.1em} &   3 \hspace{1.0em} &  1 \hspace{1.5em} \\
CaII 3934	& -7.7 \hspace{0.8em} & -12.0 \hspace{0.1em} & 123 \hspace{1.0em} & 15 \hspace{1.5em} \\
\begin{tabular}{l}
CaII 3968 \\
H$\epsilon 3970$
\end{tabular}
$\left. \begin{tabular}{l} \\ \end{tabular} \right\}$ & -8.8 \hspace{0.8em} & -22.8 \hspace{0.1em} & 145 \hspace{1.0em} & 27 \hspace{1.5em} \\
HeI 4026	& -0.4 \hspace{0.8em} &  -1.3 \hspace{0.1em} &   7 \hspace{1.0em} &  2 \hspace{1.5em} \\
H$\delta$ 4102	& -4.5 \hspace{0.8em} & -20.5 \hspace{0.1em} &  79 \hspace{1.0em} & 25 \hspace{1.5em} \\
SrII 4216	& -0.2 \hspace{0.8em} &  -0.4 \hspace{0.1em} &   3 \hspace{1.0em} &  5 \hspace{1.5em} \\
CaI 4227	& -0.1 \hspace{0.8em} &  -0.7 \hspace{0.1em} &   2 \hspace{1.0em} &  7 \hspace{1.5em} \\
FeII 4233	& -0.4 \hspace{0.8em} &  -0.7 \hspace{0.1em} &   7 \hspace{1.0em} &  8 \hspace{1.5em} \\
H$\gamma$ 4340	& -5.9 \hspace{0.8em} & -20.2 \hspace{0.1em} & 107 \hspace{1.0em} & 29 \hspace{1.5em} \\
\begin{tabular}{l}
HeI 4472 \\
FeII 4473
\end{tabular}
$\left. \begin{tabular}{l} \\ \end{tabular} \right\}$ & -0.4 \hspace{0.8em} &  -1.7 \hspace{0.1em} &   8 \hspace{1.0em} &  3 \hspace{1.5em} \\
HeII 4686	& -0.3 \hspace{0.8em} &  -0.9 \hspace{0.1em} &   6 \hspace{1.0em} &  2 \hspace{1.5em} \\
H$\beta$ 4861	& -9.4 \hspace{0.8em} & -16.8 \hspace{0.1em} & 196 \hspace{1.0em} & 30 \hspace{1.5em} \\
FeII 4924	& --- \hspace{1.0em}  &  -1.5 \hspace{0.1em} & --- \hspace{1.1em} &  2 \hspace{1.5em} \\[1ex] \hline
    \end{tabular}
  \end{center}
} 
\end{table}
The presence of inverse P Cygni profiles in the strongest emission lines
during outburst, as first noted by Herbig (1950), is here 
corroborated. At  Balmer lines higher than H9 the equivalent width of the
redshifted absorption dip is even larger than the width of the emission
component. Comparing the sequence
of spectra between March 3 and 6 we can see a substantial fading of the
absorption. The mean velocity displacement of the absorption measured in
these spectra is $+240 \pm 20$
km/s. This absorption component is still visible in our spectrum taken on
August 18 (Fig.5a). We also plot the difference between the two
spectra from August 18 and August 16 to enhance the visibility of the
absorption dip and to
remove possible contamination due to photospheric lines.
The displacement of the absorption dip measured
in the post-outburst difference spectrum corresponds to a velocity of
$+360 \pm 20$ km/s.
\begin{figure}[t]
  \centerline{\psfig{figure=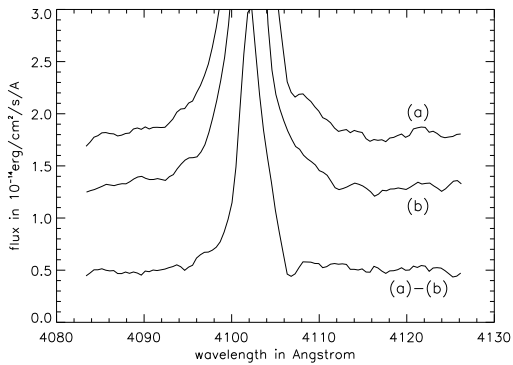,height=5.2cm}}		
  \caption[ ]{Visibility of inverse P Cygni profiles in medium resolution 
post-outburst spectra at H$\delta$. \ \
{\bf (a)} spectrum 18.08.94 \ \ {\bf (b)} spectrum 16.08.94 \ \ {\bf (a)--(b)} post-outburst variability (difference spectrum), the redshifted absorption dip
of the H$\delta$ line indicating an infall velocity of about 360km/s is clearly
visible}
\end{figure}

\subsection{Photospheric absorption features}
Photospheric features of the underlying T Tauri star can be seen only in the
post-outburst spectra. Figure 6 shows the region around CaI 4227, which is the 
strongest stellar absorption line, in two post-outburst spectra. The difference
of these two spectra no longer exhibits the absorption line, and the change of
total flux by about 40\% is therefore due to continuum emission rather than
photospheric variability.\\
The photospheric lines of the T Tauri star are veiled, even at minimum
brightness. The superimposed emission line spectrum additionally fills in
many absorption lines. The measurement of the veiling is therefore
rather difficult. We find a good fit to the observed strength of absorption
lines by introducing a flat continuum emission equal to the photospheric
continuum of the underlying star (veiling r=1, comparison with HD~202560,
spectral type M0V) at 4200 {\AA} when the
brightness of EX Lup is V=13.0.


\section{Discussion and conclusions}

The outburst of EX Lup can be understood in terms of a mass accretion event
causing increased continuum emission in a hot region close to the surface of
the star where the infalling matter finally releases its kinetic energy. The
total
light of the photosphere and the hot region becomes
dominated by the latter and therefore it is much bluer during the
outburst. Furthermore all photospheric lines are heavily veiled
(assuming that r=1 at minimum light then the veiling during outburst would be
r$\approx$20). The different slope of the continuum emission in the outburst
compared to the post-outburst (see Fig.4) indicates that the hot region
is {\it cooler} during outburst (assuming no change in extinction due
to circumstellar matter). This interpretation then implies a dramatic expansion
of the hot region in order to account for the observed rise in
luminosity during the outburst.\\
The inverse P Cygni profiles of many emission lines prove the infall motion
of accreted material. The velocity derived from the redward
displacement of the absorption component of these lines are of the order of
300 km/s and therefore much higher than those assumed in the classical boundary
layer model for T Tauri stars (Lynden-Bell \& Pringle, 1974). However, these
high infall velocities may result from magnetospherically
mediated disk accretion (Camenzind 1990, K\"onigl 1991, Hartmann et al. 1994).
High
resolution studies of classical T Tauri stars have revealed a large fraction
of stars exhibiting inverse P Cygni structures (e.g. Appenzeller 1977, Edwards
et al., 1994). The usual low level variability might be caused by geometrical
effects during the rotation of the star. The more dramatic outbursts 
could be attributed to episodic changes in the magnetosphere, resulting
in more extended infall flows of circumstellar material onto the star.\\[2ex]
{\em Acknowledgements:\,} We thank G.Herbig for alerting us to the outburst of
EX Lup in early March 1994. Also we are grateful to the following observers
for kindly providing part of their observing time: T.Abbott,
J.F.Claeskens, D.De~Winter, C.Flynn, H.Jerjen, A.Manchado, F.Patat, N.Robichon,
P.Stein.
TL \& WB were supported by student fellowships of the European Southern
Observatory. WB acknowledges support by the Deutsche Forschungsgemeinschaft 
(DFG) under grant Yo 5/16-1. 

\begin{figure}[t]
  \centerline{\psfig{figure=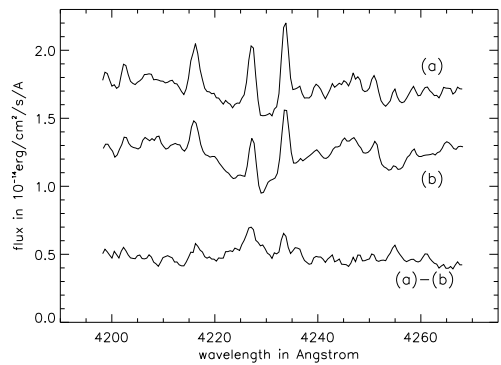,height=5.2cm}}		
  \caption[ ]{Photospheric absorption lines in medium resolution post-outburst
spectra around CaI 4227. Note also the presence of narrow emission lines SrII 
4216, CaI 4227, FeII 4233.\ \
{\bf (a)} spectrum 18.08.94 \ \ {\bf (b)} spectrum 16.08.94 \ \ {\bf (a)--(b)}
post-outburst variability (difference spectrum), photospheric absorption
disappears}
\end{figure}


\end{document}